\documentclass[aps,prl,amsmath,amssymb,preprint,superscriptaddress]{revtex4-2}

\usepackage{graphicx}
\usepackage{xcolor}
\usepackage{float}
\usepackage{bm}
\usepackage[utf8]{inputenc}
\usepackage{indentfirst}

\usepackage[unicode, colorlinks=true, linkcolor=blue, citecolor=blue, urlcolor=blue]{hyperref}

\begin{document}

\title{Entropic active particle transport in pulsating 3D geometries}



\author{Rahul Sinha} 
\email{rahulsinha@kgpian.iitkgp.ac.in}

\author{Ankit Gupta} 
\email{ankitgupta@kgpian.iitkgp.ac.in}

\author{P. S. Burada }
\email{Corresponding author: psburada@phy.iitkgp.ac.in}
\affiliation{Department of Physics, Indian Institute of Technology Kharagpur, Kharagpur 721302, India}

\date{\today}

\begin{abstract}
We study the transport of active Brownian particles (ABPs) in three-dimensional (3D) oscillatory geometries, which are spatially periodic. We establish a generalized Fick-Jacobs approach, which reduces a 3D system to an effective 1D system based on the assumption that a fast equilibration of particles along the transversal directions of the geometry. 
The transport characteristics of ABPs are computed semi-analytically and corroborated by numerical simulations. At the optimal frequency of the geometry oscillation, particles exhibit higher average velocity $\langle v \rangle$ and effective diffusion coefficient $D_{\text{eff}}$, resembling the phenomena of stochastic resonance.
This effect is further enhanced by the self-propelled velocity of ABPs and the amplitude of geometry oscillations. These findings have significant implications for the development of micro- and nanofluidic devices with enhanced control over particle transport and precise manipulation of small-scale biomedical devices.

\end{abstract}

\maketitle

Biological systems offer rich examples of fluctuating geometries used for directed transport in various natural processes. 
For instance, cilia-lined pathways in the respiratory and reproductive systems generate rhythmic oscillations that expel mucus and pathogens from the lungs and propel egg cells through the fallopian tubes, respectively  \cite{Long2024,Marcos2012}. Similarly, peristaltic motion in the gastrointestinal tract—caused by wave-like muscular contractions—moves food and nutrients efficiently along the digestive system, resembling an oscillating tube \cite{Kumral2018}. Additionally, the pulsatile expansion and contraction of blood vessels during blood flow aid in the distribution of oxygen and nutrients throughout the body \cite{Corridon2021,Wang2018}. 
These natural systems highlight how oscillatory motion can optimize the transport of mesoscopic entities, inspiring the design of synthetic microfluidic structures that mimic such mechanisms.

\begin{figure}[hbt!]
\centering
\includegraphics[scale=0.45]{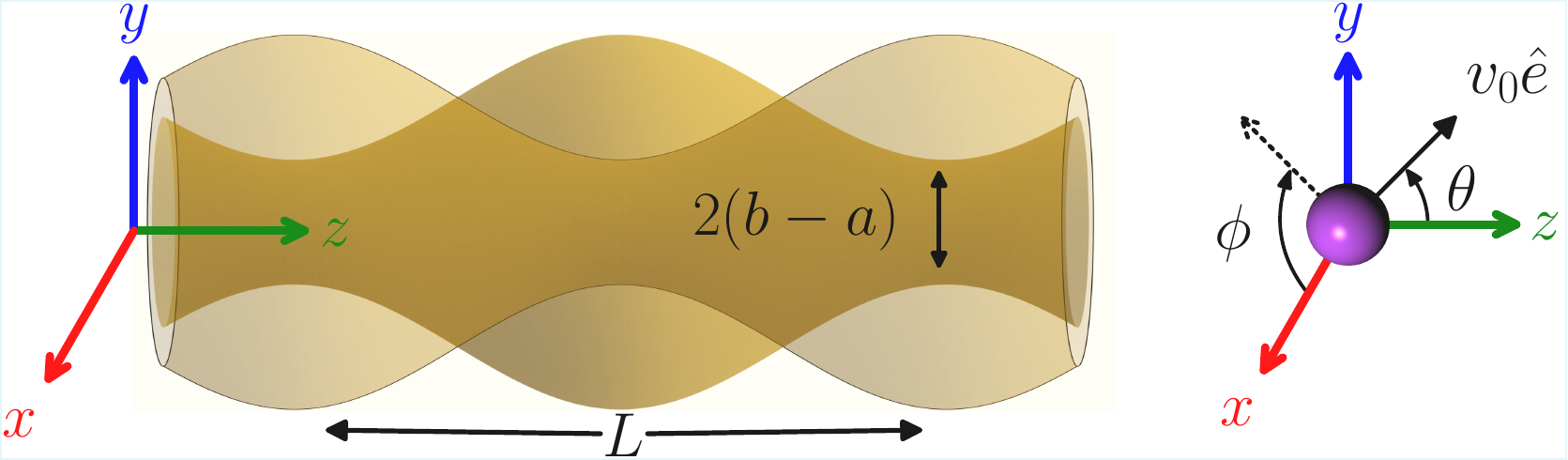}
\caption{(Left) schematic illustration of the shape a 3D oscillating tube, with time-dependent radius $W(z,t)$ (Eq.~\ref{eq:oscillations}), at time $t = 0$ (darker opacity) and at $t = T/2$ (lighter opacity).
The tube is spatially periodic along the $z$ direction with periodicity $L$ and the bottleneck diameter $2(b-a)$.
(Right) schematic of an active particle with the velocity $v_0$ and the orientation 
$\hat{e}$. $\theta$ and $\phi$ are the corresponding polar and azimuthal angles, respectively. 
}
\label{fig:model}
\end{figure}

In the scenarios mentioned above, structural agitation of the geometry can strongly influence particle transport. In these thermally or actively fluctuating 3D structures (see Fig.~\ref{fig:model}), the irregular geometry induces spatial variations along the length of the geometry, leading to the emergence of entropic barriers \cite{D_Reguera, Burada_2009, Burada2008,Gupta2023} which dictate the diffusive behavior of the particles. This effect is well captured by the Fick–Jacobs (FJ) approximation \cite{Reguera2001,Kalinay2006,Malgaretti2013}, which maps the higher-dimensional problem to an effectively one-dimensional one. 
Recent work suggests that this approach is also adapted to study fluid flow through biological membranes \cite{arango2021enhancing}, where the entropic effect significantly modifies the transport in both microfluidic \cite{Yang2017} and biological channels \cite{Rubi2017}.

A comprehensive quantitative analysis of how surface agitation influences particle transport within confined geometries remains largely unexplored. The fundamental question, whether these geometric fluctuations enhance or hinder transport, is surprisingly complex. Under certain conditions, this agitation can sometimes accelerate transport, while at other times it may impede it. Understanding the precise conditions and mechanisms governing these contrasting effects remains a challenge. Moreover, most existing studies on active particle transport focus on scenarios involving proximity to a single boundary, leaving the role of 3D confinement, where particles interact continuously with the curved geometry of enclosing surfaces.
In this letter, we study the motion of diffusing non-interacting ABPs confined in an oscillating three-dimensional symmetric tube. The oscillatory boundaries of the tube drive the particles. The main aim of this work is to study the correlation between the tube oscillations and the diffusion of ABPs using the semi-analytical approach, corroborated with the Brownian dynamics simulations.

\textit{Model -} 
Consider the dynamics of non-interacting ABPs suspended in a thermal bath and confined to diffuse within a 3D symmetric tube with time-oscillating boundaries, which is periodic in space along the $z$-direction with period $L$, as shown in Fig.~\ref{fig:model}. An ABP self-propels with a constant velocity \(v_0\) in the direction of its orientation vector $\hat{e}$,
which is expressed in terms of the polar and azimuthal angles $\theta$ and $\phi$, as shown in Fig.~\ref{fig:model}. The dynamics of ABP is governed by the overdamped Langevin equation as \cite{Vicsek1995,omar2021phase, Valani2024}

\begin{subequations}
\begin{equation}
\dot{\vec{r}}(t) = v_0 \, \hat{e}(t) + \sqrt{2D_0} \, \vec{\eta}(t),        
\end{equation}
\begin{equation}
    \dot{\hat{e}}(t) = \sqrt{2D_r} \, \vec{\xi} (t) \times \hat{e}(t).
\end{equation}
\label{eq:Langevin}
\end{subequations}
Eq.~\ref{eq:Langevin} describes the translational and rotational motion of ABP. Here, a dot over a variable denotes a derivative with respect to time, and $\vec{r}$ represents the position vector of ABP. $D_0 = k_B \Phi/\gamma_t$ and $D_r = k_B \Phi/\gamma_r$ are the translational and rotational diffusion constants that obey the Einstein relations. $k_B$ is the Boltzmann constant, $\Phi$ is the temperature of the surrounding medium, $\gamma_t$ and $\gamma_r$ denote the corresponding translational and rotational friction coefficients, respectively. $\vec{\eta}(t)$ and $\vec{\xi}(t)$ are independent, zero-mean, unit‐variance Gaussian white noises,
$ \langle \eta_i(t)\eta_j(t') \rangle = \delta_{ij} \, \delta(t-t')$ and 
$ \langle \xi(t)\xi(t') \rangle =\delta_{ij}\, \delta(t-t')$, where the indices $i, j \in \{x,y,z\}$. 
The oscillatory tube radius is defined as (see Fig.\ref{fig:model}) 
\begin{align}
W(z,t) = a \sin\left(\frac{2\pi z}{L} - \omega t\right) + b,
\label{eq:oscillations}
\end{align}
where $a$ is the amplitude of the tube oscillation, $b$ is a constant, $\omega$ is the frequency given by $\omega = 2\pi/T$, $T$ is the time period of oscillation. 
The parameters $a$ and $b$ control the curvature and diameter of the tube (see Fig.~\ref{fig:model}). To enable particle movement from one cell to the other along the length of the tube ($z-$ direction), the condition $b > a$ must hold.
Otherwise, the bottlenecks of the tube are closed.

\textit{Semi-analytical approach, Fick-Jacobs (FJ) equation -}
The Fokker-Planck equation corresponding to Eq.~\ref{eq:Langevin} is given by
\begin{align}
\frac{\partial P(\vec{r}, \hat{e}, t)}{\partial t} & = - v_0 \,\hat{e} \cdot \nabla P(\vec{r}, \hat{e}, t) \nonumber \\ 
& + D_0\, \nabla^2 P(\vec{r}, \hat{e}, t) + D_r\, \mathcal{L} P(\vec{r}, \hat{e}, t)\,,
\label{eq:Fokker}
\end{align}
where $P(\vec{r}, \hat{e}, t)$ is the probability density of ABPs \cite{vachier2019dynamics} and 
$\mathcal{L} \equiv \frac{1}{\sin\theta} \frac{\partial}{\partial \theta} 
\left( \sin\theta \frac{\partial}{\partial \theta} \right) 
+ \frac{1}{\sin^2\theta} \frac{\partial^2}{\partial \phi^2}$ is the Laplace-Beltrami operator \cite{vachier2019dynamics, sandoval2014effective}. Solving Eq.~\ref{eq:Fokker} with the no-flux boundary conditions at the tube walls is not possible analytically. However, by assuming that the diffusion time for particles along the transversal directions $(x,y)$ is much shorter compared to the diffusion time along $z$-direction, Eq.~\ref{eq:Fokker} can be integrated along $x \,\&\, y$ directions 
\cite{ Reguera2006}, and obtain an effective 1D equation analogous to the Fick-Jacobs equation for ABPs in 3D oscillatory tube. 
The 1D probability density $P(z, \hat{e}, t)$ along the $z-$ axis reads,
\begin{align}
\frac{\partial P(z, \hat{e}, t)}{\partial t} & = - v_0 \cos(\theta)\frac{\partial P(z, \hat{e}, t)}{\partial z} \nonumber \\
& + \frac{\partial}{\partial z} \Bigg[ D(z, t) e^\frac{{-A(z,t)}}{k_{B}\Phi} 
\frac{\partial}{\partial z} \Big( e^\frac{{A(z,t)}}{k_{B}\Phi} P(z, \hat{e}, t) \Big) \Bigg] \nonumber \\ 
& + D_r\, \mathcal{L} \,P(z, \hat{e}, t)\,,
\label{eq:fj_act}
\end{align}
where $\hat{e}$ is expressed in terms of $\theta$ and $\phi$.
$D(z,t)$ is the position and time-dependent diffusion coefficient incorporating both geometric curvature and activity of ABPs, defined as \cite{sandoval2014effective, modica2023boundary}
\begin{equation}
    D(z,t) = \left(D_0 + \frac{v_{0}^2}{d \,(d-1) \, D_{r}}\right) \left[ 1 + \left(h'(z,t)\right)^2\right]^{-\alpha},
\end{equation}
where $d$ is the dimension of the geometry, 
$\alpha = 1/3$ for a 2D geometry and $\alpha = 1/2$ for a 3D. The transverse cross-section is given by $h(z,t) = \pi [W(z,t)/L]^2$ in 3D and the width $h(z,t) = 2W(z,t)/L$ in 2D.  
$A(z,t)$ is the free energy of the system, which is obtained as 
\begin{equation}
    A(z,t) = E - \Phi S = - k_{B}\Phi \ln h(z,t),
\end{equation}
where $E = 0$ in our case, as there is no energetic contribution in our system, and $S = k_B \ln h(z,t)$ is the entropy due to the geometric variation in the tube's shape with respect to space and time.
In the case of passive Brownian particles (PBPs), i.e., $v_0 = 0$, the probability density is simply $P(z,t)$, recovering the standard FJ equation for PBPs \cite{Reguera2006},
\begin{equation}
\frac{\partial P (z,t)}{\partial t} = \frac{\partial}{\partial z} \left[D(z,t)e^\frac{{-A(z,t)}}{k_{B}\Phi}\frac{\partial}{\partial z} \left(e^\frac{{A(z,t)}}{k_{B}\Phi} P(z,t)\right) \right].
\end{equation}
By solving Eq.~\ref{eq:fj_act} using the initial probability density $P(z,\hat{e},0) = \delta(z-z_0) \delta(\theta-\theta_0) \delta(\phi-\phi_0)$, where $z_0, \theta_0$, and $\phi_0$ are the corresponding initial values, the translational current can be identified as 
\begin{align}
    J(z, \hat{e}, t) & = v_0 \cos(\theta) P(z,\hat{e}, t) \nonumber \\
    & - D(z,t) e^\frac{{-A(z,t)}}{k_{B}\Phi} \frac{\partial}{\partial z} \left(e^\frac{{A(z,t)}}{k_{B}\Phi}P(z,\hat{e}, t)\right)
\label{eq:current_J}
\end{align} 
The corresponding continuity equation \cite{modica2023boundary} reads, 
\begin{equation} 
\frac{\partial  P(z,\hat{e}, t)}{\partial t} + \frac{\partial J(z,\hat{e}, t)}{\partial z} = D_r \,\mathcal{L} \, P(z,\hat{e}, t). 
\end{equation}
The mean velocity $\langle v \rangle$ of ABPs  can be calculated by averaging over 
all possible orientations, space, and time.
$\langle v \rangle$ 
can be obtained as \cite{carusela2017entropic}
\begin{equation}
\langle v \rangle = \frac{1}{T} \int_0^{T} \int_{-\frac{L}{2}}^{\frac{L}{2}} \int_0^\pi \int_0^{2\pi} J(z, \theta, \phi, t) \, d\phi \, d\theta \, dz \, dt \,,
\label{eq:v_analytical}
\end{equation}
The effective diffusion coefficient $D_{\text{eff}}$ is calculated as,   
\begin{equation}
D_{\text{eff}} =  \frac{\langle z^2(t) \rangle - \langle z(t) \rangle^2}{2\,T}\,,
\label{eq:Deff_analytical}
\end{equation} 
where $\langle z \rangle$ is the mean position and 
$\langle z^2 \rangle$ is the mean-squared position of ABPs computed as
\begin{align}
\langle z(t) \rangle & = \int_{-\frac{L}{2}}^{\frac{L}{2}} \int_0^{\pi} \int_0^{2\pi}  z \,  P(z,\theta, \phi, t ) \, d\phi \,d\theta \, dz\,,\\
\langle z^2(t) \rangle & = \int_{-\frac{L}{2}}^{\frac{L}{2}} \int_0^{\pi} \int_0^{2\pi}  z^2 \,  P(z,\theta, \phi, t ) \, d\phi \,d\theta \, dz\,.
\end{align}
Note that the FJ approximation requires smooth boundary variations, i.e., $|W'(z,t)| \ll 1$. This condition ensures fast transverse equilibration, allowing the system to be effectively described by a 1D equation along the $z-$ axis, even for moderately corrugated tubes.
For simplicity, we use dimensionless quantities; we scale all lengths by the periodicity of the tube \(L\), and time by the diffusive time scale \(\tau = \gamma_t\, L^2 / k_B \Phi\). 
Now, the dimensionless variables defined as $\vec{r} \to \vec{r}/L$, $t \to t/\tau$, $v_0 \to v_0 \tau/L$, $D_0 \to D_0 \tau/L^2$, $D_r \to D_r \tau$.   
The dimensionless tube parameters are $a \to a/L$, $b \to b/L$, $\omega \to \omega \, \tau$. 
In the rest of the paper, we use dimensionless quantities.

\begin{figure}[t]
\centering
\includegraphics[scale=0.8]{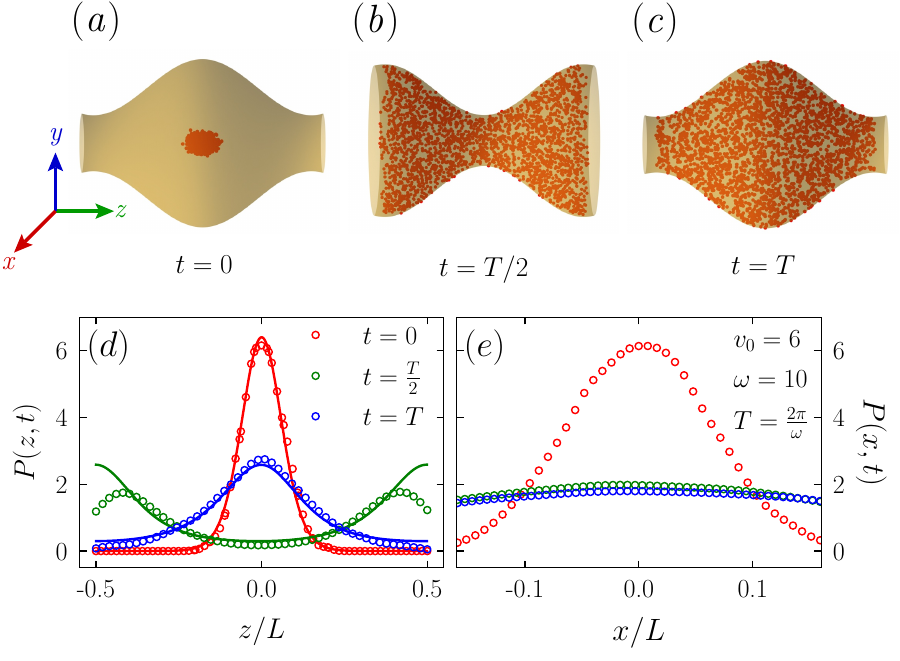}
\caption{Snapshots of active particles (with $v_0 = 6$) inside an oscillating tube at time (a) $t = 0$, (b) $t = \frac{T}{2}$ and (c) $t = T$ for $\omega = 10$. Time evolution of Probability density (d) \(P(z,t)\) along the tube axis $z$ and (e) \(P(x,t)\) (at the bottleneck) along the transversal direction of the tube $x$ at the same time intervals as snapshots were recorded. Here, we set $D_0=D_r=1$. The parameters of the tube boundary are $a = 1/2\pi $, $b = 2/2\pi$, and $L=1$}
\label{fig:pdf}
\end{figure}

\textit{Numerical simulations\, -\,}
As ABPs confined in the transversal direction of the tube, we can numerically compute $\langle v \rangle$ and $D_{\text{eff}}$
along the length of the tube ($z$-direction). 
These quantities can be calculated using the Brownian dynamics simulations performed by solving the overdamped Langevin equation (Eq.~\ref{eq:Langevin}) using the standard stochastic Euler algorithm over $2 \times 10^4$ trajectories with the integration step time $10^{-4}$. The ABPs are treated as point-sized entities initially located in the middle of the tube, in a cell [see Fig.~\ref{fig:pdf} (a) \& (d)]. When an ABP approaches the tube boundary, it essentially slides along the boundary until a thermal fluctuation in the orientational vector $\hat{e}(t)$ redirects it towards the interior of the tube \cite{Ao2014,Khatri2022,Gupta2023,Volpe2011,Gupta2024}. 
Since we considered point-sized particles, we can safely neglect the hydrodynamic interaction between the particles and the tube boundaries.
In the longtime limit, $\langle v \rangle$ and $D_{\text{eff}}$ along the $z$- direction can be calculated as
\begin{align}
\langle v \rangle & = \lim_{t \to \infty} \frac{\langle z(t) \rangle}{t}\,, 
\label{eq:V_num} \\ 
D_{\text{eff}} & = \lim_{t \to \infty} \frac{\langle z^2(t) \rangle - \langle z(t) \rangle^2}{2\,t}.
\label{eq:Deff_num}
\end{align}

\begin{figure}[thb!]
\centering
\includegraphics[scale=1.2]{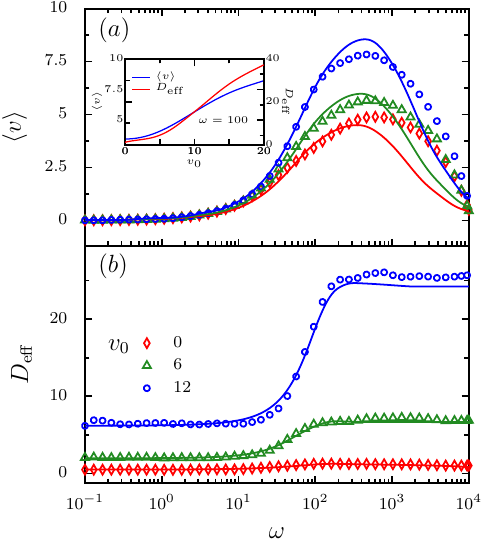}
\caption{$(a)$ The mean velocity \(\langle v \rangle\) and $(b)$ the effective diffusion coefficient \(D_{\text{eff}}\) as a function of the tube oscillation frequency \(\omega\) for various values of the self-propulsion speed \(v_0\) of the particles. 
Symbols correspond to numerical simulation results (Eqs.~\ref{eq:V_num} \,\&\, ~\ref{eq:Deff_num}), whereas solid lines indicate the semi-analytical results (Eqs.~\ref{eq:v_analytical} \& ~\ref{eq:Deff_analytical}) obtained by using the FJ approximations. The inset in panel~(a) presents the semi-analytically obtained  $\langle v \rangle$ and $D_{\text{eff}}$ as functions of $v_0$ at a fixed frequency $\omega = 100$. 
Here, we set $D_0 = D_r = 1$. The tube parameters are \(L = 1\) , \(a = 1/2\pi\) and \(b = 2/2\pi\).}
\label{fig:omega}
\end{figure}

\textit{Transport Characteristics \, - \,}
The behavior of $\langle v \rangle$ and $D_{\text{eff}}$ as a function of frequency $\omega$ of the oscillatory tube for different values of the self-propulsion speed $v_0$ is depicted in Fig.~\ref{fig:omega}.  
Surprisingly, there is a very good agreement between analytical predictions and the numerical results up to a higher value of $\omega$. 
Which means that the FJ approximation is valid even in the case of oscillatory boundaries. Of course, the boundaries should not rapidly oscillate.    
In a static tube, or a tube with weak oscillatory boundaries, $\langle v \rangle$ is nearly zero and $D_{\text{eff}}$ is finite, as they should be, because there is no external force acting on the particles.  
In this case, particles have sufficient time to explore the space available in the transversal directions ($x$ and $y$), satisfying the FJ approximation.
As $\omega$ increases, due to the movement of the tube boundaries in a particular direction, both 
$\langle v \rangle$ and $D_{\text{eff}}$ increase. Interestingly, the FJ approximation still provides accurate results. 
It is observed that the magnitude of $\langle v \rangle$ and $D_{\text{eff}}$ increases with the self-propelled velocity $v_0$ of ABPs, as the diffusive nature of ABPs increases with $v_0$, allowing particles to explore the available space inside the tube more effectively.

For intermediate values of $\omega$, particles exhibit maximum $\langle v \rangle$ and higher $D_{\text{eff}}$.
This is where the oscillatory boundaries of the tube most effectively enhance the particle movement. 
This situation resembles the Stochastic Resonance \cite{gammaitoni1998stochastic} phenomena observed in the case of confined channels \cite{Burada2008}. The FJ nearly captures these maximum values. Notably, the peak position is independent of the self-propulsion speed $v_0$.
When the $\omega$ value is further increased, particles do not have enough time to explore the space available in the transversal directions. Effectively, particles are restricted to the bottleneck regions (see Fig.~\ref{fig:omega}).
As a result, FJ approximation fails \cite{Reguera2006,Burada_2009}.

In the limit, $\omega\to\infty$, i.e., when the tube boundaries are rapidly changing with a time scale much smaller than the diffusion time scale of the particles, particles do not feel the oscillations of the tube boundaries and particle movement is restricted around the bottleneck region. This situation is analogous to the case of particles in a cylindrical tube, and as one would expect, $\langle v \rangle \to 0$ and $D_{\text{eff}}$ satisfies \cite{zottl2016emergent}, 
\begin{align}
D_{\text{eff}} =  D_0 + \frac{v_0^2}{d(d-1)D_r}\,.
\label{eq:D_lim}
\end{align}
Interestingly, FJ approximation nearly captures this result. 

The probability density of ABPs along the length of the tube $P(z,t)$ and along one of the transversal directions $P(x,t)$ are depicted in Fig.~\ref{fig:pdf}(d) \&(e), respectively.
Initially ($t = 0$), all the particles were placed in the middle of the tube [see Fig.~\ref{fig:pdf}(a) \& (d)]. 
As time progresses, the particles diffuse inside the tube, exploring all the available space, and feel the oscillations of the tube boundaries. These behaviors are nearly captured by the FJ approximation and numerical simulations. 
See Fig.~\ref{fig:pdf}(b) \& (c). 
Also, in support of the FJ approximation, particles are uniformly distributed along the transversal directions, as shown in Fig.~\ref{fig:pdf}(e). 
Note that at lower $\omega$, the diffusion time is much smaller than the oscillations of the tube, and particles get enough time to explore the available space inside the tube.

\begin{figure}[t]
\centering
\includegraphics[scale=1.2]{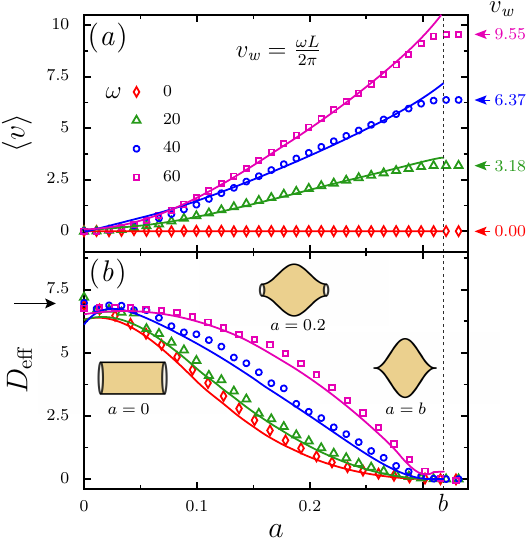}
\caption{(a) The magnitude of the mean velocity \(\langle v \rangle\). The left arrows indicate the values of saturated mean velocity \(\langle v\rangle = v_w= \frac{\omega L}{2\pi}\) beyond \(a=b\) and (b) the effective diffusion coefficient \(D_{\text{eff}}\) as a function of the corrugation amplitude $a$ for different values of boundary oscillation frequency \(\omega\). 
The right arrows denote the value of $D_{\text{eff}}$ calculated using Eq.~\ref{eq:D_lim} for a given $v_0$. 
Here, we set \(D_0=D_r= 1\), and \(L = 1\). The other parameters are \(v_0 = 6\) and $b=2/2\pi$. } 
\label{fig:epsilon}

\end{figure}

Fig. \ref{fig:epsilon} shows $\langle v \rangle$ and  $D_{\text{eff}}$ as a function of the amplitude of the tube oscillation $a$ (see Eq.~\ref{eq:oscillations}) for different values of frequency $\omega$. FJ approximation results are in good agreement with the numerical simulation results. 
When $a = 0$, the tube is a cylinder with radius $b$, and as expected, $\langle v \rangle$ is zero and $D_{\text{eff}}$ satisfies Eq.~\ref{eq:D_lim}.
However, in the case of a static and modulated tube, i.e., $\omega = 0$ and $a \neq 0$, due to symmetry of the tube $\langle v \rangle$ is zero and $D_{\text{eff}}$ decays monotonically as $a$ increases.
The tube expands in the transversal directions with increasing $a$, and as a result, particles explore the available space as they diffuse, which leads to a decrease in $D_{\text{eff}}$ along the tube axis. In this situation, the entropic barriers play a vital role \cite{Reguera2006,Burada_Chemphy} (see Fig.~\ref{fig:epsilon}).  
As $\omega$ increases, the tube boundaries squeeze the particles along the $z-$direction. As a result, both $\langle v \rangle$ and $D_{\text{eff}}$ increase.
When $a \approx b$, $\langle v \rangle$ saturates at the phase velocity of the tube boundaries ($v_{\text{wall}} = \frac{\omega L}{2\pi}$). At $a = b$, i.e, when the pore radius ($b-a = 0$), the bottle-necks of the tube are closed and particles are confined without moving from one cell to another. 
Particles move together as a wave packet, and $\langle v \rangle$ matches the phase velocity of the tube walls. 
This confinement leads to vanishing $D_{\text{eff}}$ as the particles are unable to spread between the adjacent cells of the tube. 
In this situation, the FJ approximation fails as there is no accessible transverse domain, preventing equilibration in the transverse direction.  Consequently, both the free energy $A(z,t)$ and the translational current $J(z,t)$ diverge.

\textit{Discussion\, - \,}
Particle diffusion confined within the fluctuating boundaries or dynamic porous media shows transport behavior distinct from the case of static confinement \cite{wang2023interactions,sarfati2021enhanced, nagella2023colloidal}. 
For example, our semi-analytical and numerical simulations reveal a resonance-like enhancement \cite{romanczuk2010quasideterministic, torrenegra2022enhancing} in both $\langle v \rangle$ and $D_{\text{eff}}$ of ABPs. In contrast with the intuition that transport increases (decreases) with high (low) oscillation frequency $\omega$ due to structural jiggling, we find that enhancement peaks at an optimal frequency.
This observation is consistent with the experimental results of thin liquid films and oscillating microchannels showing frequency-dependent mobility enhances \cite{Stubenrauch2003}. Additionally, studies on particles and macromolecules confined between oscillating surfaces report increased mobility and diffusion at a certain frequency, supporting our findings \cite{marbach2018transport, Bleil2007}. These parallels between semi-analytics, simulations, and experiments underscore the physical relevance of resonance-enhanced transport in confined systems. 
To contextualize our dimensionless parameters in real units consider, particles moving in a 3D oscillating geometry with system parameters $a \sim 15.9  \, \mu m$, $b \sim 31.8 \, \mu m$ and period length $L \sim 100 \, \mu m$ filled with water at room temperature ($\Phi \sim 300 \, K$ ), having viscous drag $\gamma_t \sim 2 \times 10^{-10} kg/s$, characteristic diffusion time $\tau \sim 455 \,s$, and self propulsion speed $v_0 \sim 1.3 \, \mu m/s$, which corresponds to the dimensionless propulsion speed $v_0 = 6$ used in our model, our dimensionless frequency range $\omega = 0.1\text{---}10000$ corresponds to a physical band of $\approx 3.5 \times 10^{-5}\text{---}3.5\,\, Hz$.

\textit{Conclusions\, - \,}
In conclusion, this work highlights the crucial role of confinement and the coupling between the intrinsic activity of the particles and the dynamic oscillations of the tube boundaries. This coupling gives rise to novel and non-trivial transport phenomena that are not accessible in static systems, highlighting the universality of resonance-like mechanisms in particle transport. Our model more accurately captures natural and synthetic systems with confinement in 3D. We have demonstrated that the transport properties have maximum values at an intermediate frequency $\omega$ of the tube oscillations,
maximizing both  $\langle v \rangle$ and $D_{\text{eff}}$ of ABPs. Furthermore, the amplitude of boundary oscillations enhances both $\langle v \rangle$ and $D_{\text{eff}}$. In the limit of tube bottleneck radius to zero, $\langle v \rangle$ approaches the phase velocity of the oscillating wall ($v_{\text{wall}} = \frac{\omega L}{2\pi}$) and $D_{\text{eff}}$ is nearly zero, where particles are confined to individual cells. Our framework reveals key principles for controlling particle transport in confined systems with oscillatory boundaries. In the latter case, one needs to identify the appropriate length scale depending on the system of interest. These results suggest improved strategies for targeted drag delivery, microfluidic control, selective filtration, and precise biomolecular manipulation, including the regulation of DNA and RNA during sequencing \cite{Kowalczyk2012}.

This work was financially supported by the University Grants Commission (UGC) 
India fellowship in science and SERB, Government of India, project-wide CRG/2023/006186.
The authors acknowledge the Indian Institute of Technology (IIT) Kharagpur for providing research facilities.

\bibliographystyle{apsrev4-2}

\bibliography{arxiv}

\end{document}